\def\year{2015}
\documentclass[letterpaper]{article}
\usepackage{aaai}
\usepackage{times}
\usepackage{helvet}
\usepackage{courier}

\usepackage[lowtilde]{url} 
\usepackage{graphics}
\usepackage{graphicx}
\usepackage{amsmath}
\usepackage{rotating}
\usepackage{subfig}

\begin{document}
%
\title{An Image is Worth More than a Thousand Favorites:\\Surfacing the Hidden Beauty of Flickr Pictures}
\author{Rossano Schifanella\\University of Turin\\Turin, IT\\schifane@di.unito.it \And Miriam Redi\\ Yahoo Labs\\Barcelona, SP\\redi@yahoo-inc.com \And Luca Maria Aiello\\Yahoo Labs\\Barcelona, SP\\alucca@yahoo-inc.com} 

\maketitle
\begin{abstract}
\begin{quote}
The dynamics of attention in social media tend to obey power laws. Attention concentrates on a relatively small number of popular items and neglecting the vast majority of content produced by the crowd. Although popularity can be an indication of the perceived value of an item within its community, previous research has hinted to the fact that popularity is distinct from intrinsic quality. As a result, content with low visibility but high quality lurks in the tail of the popularity distribution. This phenomenon can be particularly evident in the case of photo-sharing communities, where valuable photographers who are not highly engaged in online social interactions contribute with high-quality pictures that remain unseen. We propose to use a computer vision method to surface beautiful pictures from the immense pool of near-zero-popularity items, and we test it on a large dataset of creative-commons photos on Flickr. By gathering a large crowdsourced ground truth of aesthetics scores for Flickr images, we show that our method retrieves photos whose median perceived beauty score is equal to the most popular ones, and whose average is lower by only 1.5\%.
\end{quote}
\end{abstract}

\section{Introduction}\label{sec:intro}

One of the common uses of online social media surely is to accrue social capital by winning other people's attention~\cite{steinfield08social,smith10bonding,burke11differentiating,bohn14making}. The ever-increasing amount of content produced by the crowd triggers emergent complex dynamics in which different pieces of information have to compete for the limited attention of the audience~\cite{romero11influence}. In this process, only few individuals and the content they produce emerge and become popular, while the vast majority of people are bound to a very limited visibility, their contributions being rapidly forgotten~\cite{cha07tube,sastry12head}. Such dynamics do not necessarily promote high-quality content~\cite{weng12competition}, possibly confining some valuable information and expert users in the very tail of the popularity distribution~\cite{goel10anatomy}. This might cause a loss to the community, first because tail contributors are likely to lose engagement and churn out~\cite{karnstedt11effect}, but also because tail content is often less curated and difficult to find through search~\cite{baeza13online}.

Previous work has focused extensively on studying the patterns of popularity of social media users and of all sorts of online content, trying to isolate the predictive factors of success~\cite{suh10want,hong11predicting,brodersen12youtube,khosla14image}. However, considerably less effort has been spent in finding effective ways to surface high-quality content from the sea of forgetfulness of the popularity tail. Finding valuable content in the pool of unpopular items is an intrinsically difficult task because tail items \textit{i)} are many, outnumbering by orders of magnitude those with medium or high popularity, \textit{ii)} have most often low quality, making random sampling strategies substantially ineffective, and \textit{iii)} tend to be less annotated and therefore more difficult to index.

We contribute to tackle these problems in the context of photo sharing services. We use a computer vision method to surface beautiful pictures among those with near-zero-popularity, with no need of additional metadata. Our approach is supervised and relies on features developed in the field of computational aesthetics~\cite{datta}. To train our framework, we collect for the first time a large ground truth of aesthetic scores assigned to Flickr images by non-expert subjects via crowdsourcing. Differently from conventional aesthetics datasets~\cite{datta,murray2012ava}, our ground truth includes images with a wide spectrum of quality levels and better reflects the taste of a non-professional public, making it the ideal training set to classify web images.

When tested on nearly 9M creative-commons Flickr pictures, our method is able to surface from the set of photos that received very low attention ($\leq$5 favorites) a selection of images whose perceived beauty is close to that of the most favorited ones, with the same median value and an average value that is just $1.5\%$ lower. Results are consistent for images in four different topical categories and largely outperform a random baseline, computer vision methods trained on traditional aesthetics databases, and a state-of-the-art computer vision methods targeted to the prediction of image popularity~\cite{khosla14image}.

\noindent We summarize our main contributions as follows:
\begin{itemize}
	\item We build and make publicly available\footnote{\scriptsize\url{http://di.unito.it/beautyicwsm15}} the largest ground truth of aesthetic scores for Flickr photos constructed so far, including 10.800 pictures of 4 different topical categories and 60K judgments. We carefully designed the crowdsourcing experiment to account for the biases that can incur in a task that is characterized by a strong subjective component.
	\item We provide an analysis of ordinary people's aesthetics perception of web images. We find that perceived beauty and popularity are correlated ($\rho=0.43$) but the beauty scores of very popular items have higher variance than unpopular ones. We find that a non-negligible number of unpopular items are extraordinarily appealing. 
	\item We propose a method to retrieve beautiful yet unpopular images from very large photo collections. Our approach works in a pure cold start scenario as it needs in input only the visual information of the picture. Also, it overcomes the issue of sparsity (i.e., very few beautiful pictures hidden among very large amounts of mediocre images) with surprisingly high precision, being able to retrieve images whose perceived beauty is comparable to the top-rated photos.
\end{itemize}
After a review of the related work ($\S$\ref{sec:related}), we touch upon the popularity skew in Flickr ($\S$\ref{sec:popularity}). We then describe the process of collection of the aesthetics scores through crowdsourcing ($\S$\ref{sec:crowdflower}). Next, we describe the computer vision method we use to identify beautiful pictures ($\S$\ref{sec:aesthetics}) and we report the aesthetic prediction results in comparison with other baselines ($\S$\ref{sec:prediction}). Last, we show that our method can surface beautiful photos from a large pool of non-popular ones ($\S$\ref{sec:surfacing}).

\section{Related work}\label{sec:related}

\noindent \textbf{Popularity Prediction.} Being able to characterize and predict item popularity in social media is an important, yet not fully solved task~\cite{hong11predicting}. The possibility of predicting the popularity of videos and pictures in social platforms like YouTube, Vimeo, and Flickr has been explored extensively~\cite{cha09measurement,figueiredo11tube,brodersen12youtube,ahmed13peek}. Multimodal supervised approaches that combine metadata and computer vision features have been used to predict photo popularity. Visual features like coarseness and colorfulness, well predict the number of favorites in Flickr~\cite{sanpedro09ranking} and the number of reshares in Pinterest to some extent~\cite{totti14impact}. The presence of specific visual concepts in the image, such as human faces~\cite{bakhshi14faces}, are good predictors too. Recently, Khosla et al.~\cite{khosla14image} have made one of the most mature contributions in this area, training a SVR model on both visual content and social cues to predict the normalized view count on a large corpus of Flickr images. While previous work tries to understand why popular images are successful, we flip the perspective to see if high-quality pictures hide in the long tail and to what extent we are able to automatically surface them. This necessity is also supported by the weak correlation between received attention and perceived quality found in small image datasets~\cite{liang14investigating}.

\vspace{4pt} \noindent \textbf{Popularity vs. Quality.}
Both social and computer scientist have investigated the relation between popularity and intrinsic quality of content. Items' popularity is only partly determined by their quality and it is largely steered by the early popularity distribution, often with unpredictable patterns~\cite{salganik2006experimental}. User's limited attention drives the popularity persistence and virality of an item more than its intrinsic appeal~\cite{weng12competition,hodas12visibility}. A piece of content can attract attention because of many factors including the favorable structural position of its creator in a social network~\cite{hong11predicting}, the sentiment conveyed by the message~\cite{quercia12mood}, or the demographic~\cite{suh10want} and geographic~\cite{brodersen12youtube} composition of the audience. On video~\cite{sastry12head} or image~\cite{zhong13sharing} sharing platforms, the content that receives larger shares of attention is often of niche topical interest. Adopting community-specific behavioural norms can also increase popularity returns. On Twitter, users who generate viral posts are those who limit their tweets to a single topic~\cite{cha10measuring}. On Facebook, communicating along weak ties is the key to spread content~\cite{bakshy12role}. More in general, social activity, even in its most superficial meaning (e.g., ``poking'') can be a powerful attractor of popularity~\cite{epjds14vaca,aiello12icwsm}.

\vspace{4pt} \noindent \textbf{Computational Aesthetics.}
Computational aesthetics is the branch of computer vision that studies how to automatically score images in terms of their photographic beauty. Datta et al.~(\citeyear{datta}) and Ke et al.~(\citeyear{ke2006design}) designed the first compositional features to distinguish amateur from professional photos. Computational aesthetics researchers have been developing dedicated discriminative visual features and attributes~\cite{nishiyama2011aesthetic,dhar11high}, generic semantic features~\cite{marchesotti2011assessing,murray2012ava}, topic-specific models~\cite{luo08photo,obrador09role} and effective learning frameworks~\cite{wu2011learning} to improve the quality of the aesthetics predictors. Aesthetic features have been also used to infer higher-level properties of images and videos, such as image affective value~\cite{emotions}, image memorability~\cite{isola2011}, video creativity~\cite{redi6}, and video interestingness~\cite{redi12where,jiang2013understanding}. To our knowledge, this is the first time that image aesthetic predictors are used to expose high quality content from low-popular images in the context of social media.

\vspace{4pt} \noindent \textbf{Ground Truth for Image Aesthetics.} Existing aesthetic ground truths are often derived from photo contest websites, such as DPChallenge.com~\cite{ke2006design} or Photo.net~\cite{datta}, where (semi) professional photographers can rate the quality of their peers' images. The average quality and style of the images in such datasets is way higher than the typical picture quality in photo sharing sites, making them not suitable to train \textit{general} aesthetic models. Hybrid datasets~\cite{luo2011content} that add lower-quality images to overcome this issue are also not good for training~\cite{murray2012ava}. In addition, social signals such as Flickr interestingness\footnote{\scriptsize Flickr interestingness algorithm is secret, but it considers some metrics of social feedback. For more details refer to \url{https://www.flickr.com/explore/interesting}}~\cite{jiang2013understanding} are often used as a proxy for aesthetics in that type of datasets. However, no quantitative evidence is given that neither the Flickr interestingness nor the popularity of the photographers are good proxies for image quality, which is exactly the research question we address. Crowdsourcing constitutes a reliable way to collect ground truths on image features~\cite{Redi:2014}, the only attempt to do it in the context of aesthetics has been limited in scope (faces) and very small-scale~\cite{li2010aesthetic}.



\section{Popularity in Flickr}\label{sec:popularity}

Flickr is a popular social platform for image sharing. Users can establish directed \textit{social links} by ``following'' other users to get updates on their activity. Users can label their own photos with free-text \textit{tags} and publish them in the photo pools of \textit{groups}. Every public photo can be marked as \textit{favorite} or annotated with a textual \textit{comment} by any user in the platform. Flickr also maintains and updates periodically the \textit{Explore} page\footnote{\scriptsize \url{https://www.flickr.com/explore}}, a showcase of interesting photos.

\begin{figure}[t!]
    \includegraphics[width=0.48\textwidth]{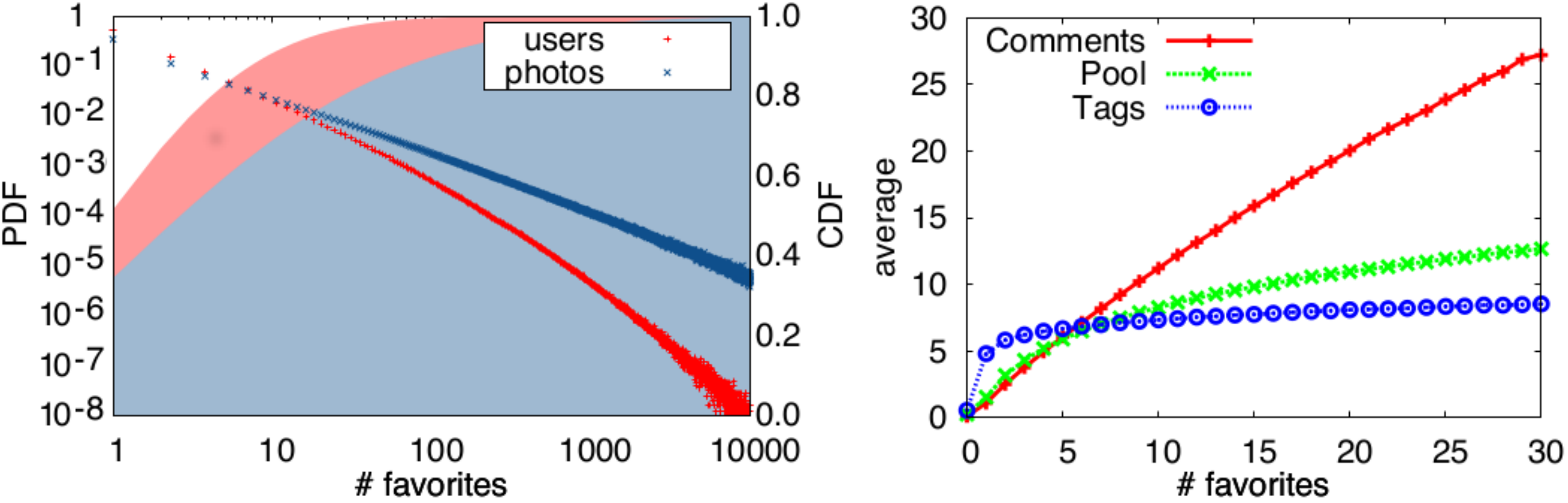}

   \caption{\small (Left) Distribution of the number of favorites for Flickr photos and users. (Right) Average number of comments, tags, and uploads to group photo pools for photos with a fixed number of favorites.}
	\vspace{-1.5em}
    \label{fig:popularity}
  \end{figure}

The complex dynamics that attract attention towards Flickr images revolve around all the above mentioned mechanisms of social feedback that, as in any other social network, tend to promote some items more than others. As a result, the distribution of picture popularity ---usually measured by the number of favorites~\cite{cha09measurement}--- is very broad. Figure~\ref{fig:popularity} (left) shows statistics on user and image popularity computed over a random sample of 200M public Flickr photos that have been favorited at least once. The distribution of the mass of favorites over the photos is highly unequal (Gini coefficient $0.68$): the number of favorites of the pictures in this sample spans four orders of magnitude, with the majority of them having only one favorite (52\%). The same figure holds when aggregating the popularity by users: some accumulate thousands favorites while the vast majority ($\sim$70\%) rustles up less than ten.

As for the intuition given by the \textit{Infinite Monkey Theorem}, the unpopular users must be able to collectively produce a certain amount of exceptionally valuable content just because of their substantial number. More concretely, it is hard to believe that there is no high-quality photo among 166M pictures with five favorites or less. Estimating how many beautiful pictures lie in the popularity tail and understanding how we can draw those out of the immense mass of user-generated content are the main goals of this contribution.

One may think that one possibility to achieve the goal would be to leverage different types of social feedback (e.g., comment). However, unpopular items rarely receive social feedback. As displayed in Figure~\ref{fig:popularity} (right), the number of comments, tags, and uploads in groups is positively correlated with the number of favorites, with near-zero favorite pictures having a near-zero amount of all the other metrics, on average. Providing a method that does not rely on any type of explicit feedback has therefore the advantage of being more general and suitable for a cold-start scenario. For this reason, we rely on a supervised computer vision method that we describe in $\S$\ref{sec:aesthetics} and whose training set is collected as described in the next section.

\section{Ground truth for image aesthetics} \label{sec:crowdflower}

We build a ground truth for aesthetics from a 9M random sample of the Creative Commons Flickr Images dataset\footnote{ \scriptsize \url{http://bit.ly/yfcc100m}}. We collect the annotations
using CrowdFlower\footnote{\scriptsize \url{http://www.crowdflower.com}}, a large crowdsourcing platform that distributes small, discrete \textit{tasks} to online \textit{contributors}. Next we describe how we selected the images for our corpus ($\S$\ref{sec:crowdflower:dataset}), how we run the crowdsource experiment ($\S$\ref{sec:crowdflower:experiment}), and the results on the beauty judgments we got from it ($\S$\ref{sec:crowdflower:results}).

\begin{table}[t!]
\centering
\small
\begin{tabular}{|c|p{6cm}|} \hline
\textbf{Category} & \multicolumn{1}{c|}{\textbf{Tags}} \\ \hline
\textbf{people} & people, face, portrait, groupshot  \\
\textbf{nature} & flower, plant, tree, grass, meadow, mountain \\
\textbf{animals} & animal, insect, pet, canine, carnivore, butterfly, feline, bird, dog, peacock, bee, lion, cat\\
\textbf{urban} & building, architecture, street, house, city, church, ceiling, cityscape, brick, tower,  window, highway, bridge \\
\hline
\end{tabular}
\caption{\small  Set of machine tags included in each image category}
\vspace{-1.5em}
\label{tab:cat2tags}
\end{table}

\subsection{Definition of the image corpus} \label{sec:crowdflower:dataset}

To help the contributor in the assessment of the image beauty, we build a photo collection that \textit{i)} presents topically coherent images and \textit{ii)} represents the full popularity spectrum, thus ensuring a diverse range of aesthetic values. 

\vspace{4pt} \noindent \textbf{Topical Coherence.} Different picture categories can achieve the same aesthetic quality driven by different criteria~\cite{luo2011content}. To make sure that contributors use the same evaluation standard, we group the images in classes of coherent subject \textit{categories}. To do that, we use Flickr \textit{machine tags}\footnote{\scriptsize \url{http://bit.ly/1umsOnL}}, namely tags assigned by a computer vision classifier trained to recognize the type of subject depicted in a photo (e.g., a bird or a tree) with a certain confidence level.  We  manually group the most frequent machine tags in topically-coherent macro-groups, coming up with 4 final categories: \textit{people}, \textit{nature}, \textit{animals}, and \textit{urban}. We only consider the pictures associated with high-confidence machine tags ($\geq$0.9). Moreover, we manually clean the final photo selection by replacing few instances that suffered from machine tag misclassification. The full list of machine tags per category is reported in Table~\ref{tab:cat2tags}.

\vspace{4pt} \noindent \textbf{Full Popularity Range.} Within each category, we are interested in assessing the perceived beauty of photos with different popularity levels. To do so, we identify three popularity buckets obtained by logarithmic binning over the range of number of favorites $f$. We refer to them as \textit{tail} ($f\leq5$), \textit{torso ($5<f\leq45$)}, and \textit{head} ($f>45$). The tail of the distribution contains $98\%$ of the photos, whereas the torso and head contain $1.6\%$ and $0.4\%$ respectively. We randomly sample, within each category, 1000 images from the \textit{tail} and 1000 from the \textit{torso}. Because of the reduced number of most popular pictures we do not sample randomly the \textit{head} bucket but we consider the top 500 instead. Images from such diverse popularity levels are also likely to take a wide range of aesthetic values, thus ensuring aesthetic diversity in our corpus, very important to get reliable beauty judgements~\cite{redi2013crowdsourcing}.

\subsection{CrowdFlower experiment}\label{sec:crowdflower:experiment}

Crowdsourcing tasks are influenced by a variety of human factors that are not always easy to control~\cite{mason14conducting}. However, platforms like CrowdFlower offer advanced mechanisms to tune the annotation process and enable the best conditions to get high-quality judgments. To facilitate the reproducibility of our experiment, next we report the main setup parameters.

\begin{figure}[t!]
\begin{center}
\includegraphics[width=0.99\columnwidth]{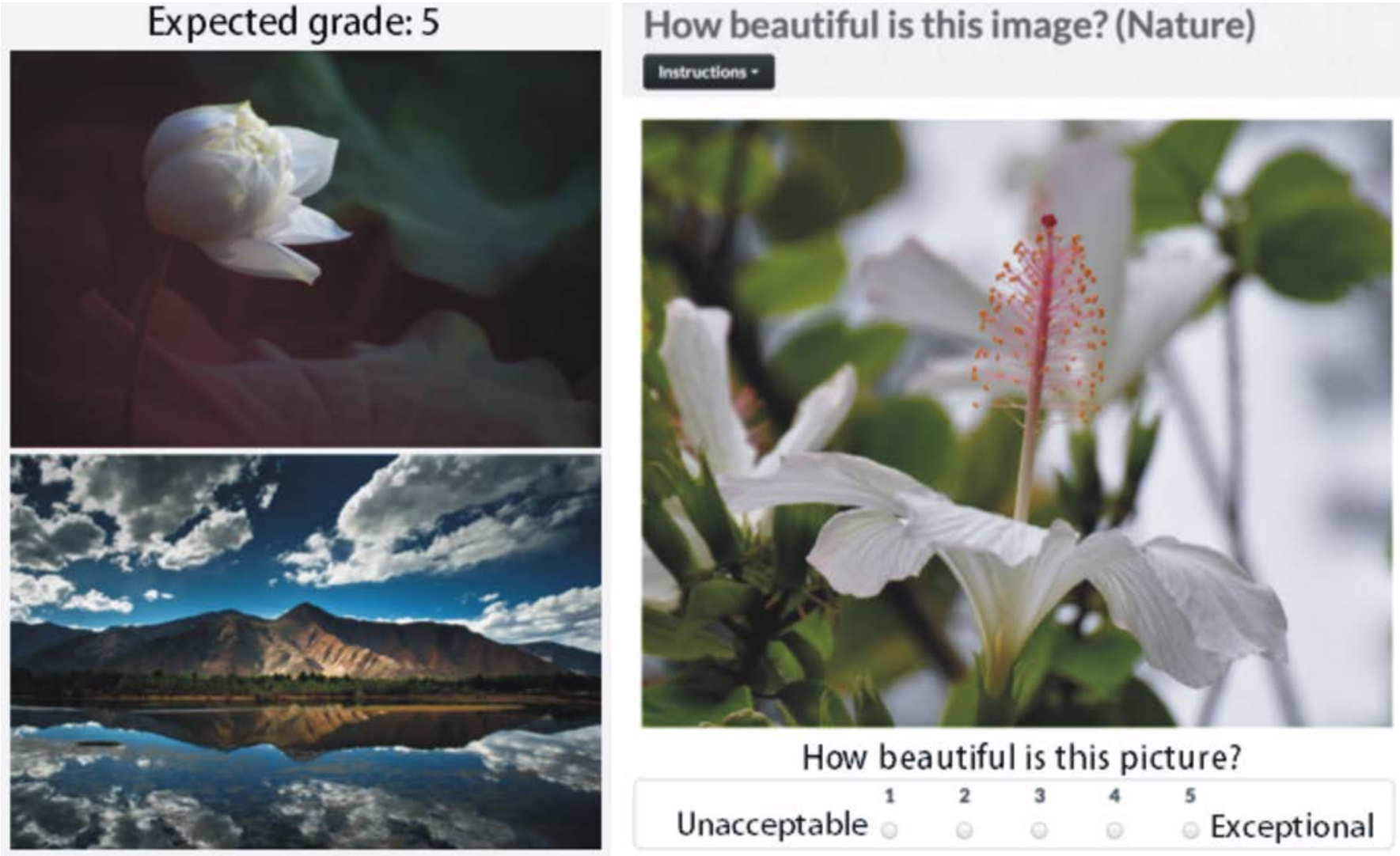}
\caption{\small  Screenshot of the crowdflower job: instruction examples (left) and voting task (right).}
\vspace{-1.5em}
\label{fig:screenshotz}
\end{center}
\end{figure}

\vspace{4pt} \noindent \textbf{Task interface and setup}. The task consists in looking at a number of images and evaluating their aesthetic quality. At the top of the page we report a short description of the task and we ask \textit{``How beautiful is this picture?''}. The contributor is invited to judge the intrinsic beauty of an image and not the appeal of its subject; high quality, artistic pictures that depict a non-conventionally beautiful subject (e.g., a spider), should be marked as beautiful and viceversa. Screenshots of the Crowdflower job interface are shown in Figure~\ref{fig:screenshotz}.

Although several approaches and rating scales can be used to get quality feedback~\cite{fleet14interestingness}, we use the 5-point \textit{Absolute Category Rating} (ACR) scale, ranked from \textit{``Unacceptable''} to \textit{``Exceptional''}, as it is a good way to collect aesthetic preferences~\cite{redi2014beauty}. To help the annotators in their assessment, two example images and a textual description of each grade are provided (see Figure~\ref{fig:screenshotz} and Table~\ref{tab:likertscale}). The examples are Flickr images that have been unanimously judged by three independent annotators to be clear representatives of that beauty grade. Below the examples, each page contains 5 randomly selected images (\textit{units} of work in CrowdFlower jargon), each followed by the radio buttons to cast the vote. The random selection of images allows us to mix pictures from different popularity ranges in the same page, thus offering to the users an easier context for comparison~\cite{fleet14interestingness}. We show all the images with approximately the same (large) size because image size can skew the perception of image quality~\cite{chu13size}.

Each photo receives at least 5 judgments, each one by a different independent contributor. Each contributor can submit a maximum of 500 judgments, to prevent a predominance of a small group of workers. Contributors are geographically limited to a set of specific countries\footnote{\scriptsize Australia, Austria, Belgium, Denmark, Finland, France, Germany, Ireland, Italy, Netherlands, Poland, Spain, Sweden, United Kingdom, United States}, to ensure higher cultural homogeneity in the assessment of image aesthetics~\cite{hagen78perception}. Only contributors with an excellent track record on the platform (responsible for the $7\%$ of monthly CrowdFlower judgments overall) have been allowed. We also banned workers that come from external crowdsourcing channels that have a ratio of trusted/untrusted users lower than $0.9$.

\begin{table}[t!]
\begin{center}
\small
\begin{tabular}{|c|c|p{5cm}|}
\hline
1 & \textbf{Unacceptable} & Extremely low quality, out of focus, underexposed, badly framed images   \\ 
2 & \textbf{Flawed} & Low quality images with some technical flaws (slightly blurred, slightly over/underexposed, incorrectly framed) and without any artistic value \\
3 & \textbf{Ordinary} & Standard quality images without technical flaws (subject well framed, in focus, and easily recognizable) and without any artistic value \\
4 & \textbf{Professional} & Professional-quality images (flawless framing, focus, and lightning) or with some artistic value \\
5 & \textbf{Exceptional} & Very appealing images, showing both outstanding professional quality (photographic and/or editing \& techniques) and high artistic value \\
\hline
\end{tabular}
\end{center}
\vspace{-10pt}
\caption{\small Description of the five-level aesthetic judgment scale}
\vspace{-1.5em}
\label{tab:likertscale}
\end{table}

\vspace{4pt} \noindent \textbf{Quality control.} \textit{Test Questions} (also called \textit{Gold Standard}) are used to test and track the contributor's performance and filter out bots or unreliable contributors. To access the task, workers are first asked to annotate correctly 6 out of 8 \textit{Test Questions} in an initial \textit{Quiz Mode} screen and their performance is tracked throughout the task with \textit{Test Questions} randomly inserted in every task, disguised as normal units. To support the learning process of a contributor, we tag each \textit{Test Question} with an explanation that pops up in case of misjudgment (e.g., ``excellent combination of framing, lightning, and colors resulting in an artistic image, visually very appealing'' is one of the description for an high rated item).

To build the set of \textit{Test Questions}, we first collected about 200 candidate images from different online sources including Flickr, web repositories, aesthetics corpora~\cite{murray2012ava}, and relevant photos retrieved by the main image search engines. Three independent editors manually annotated the candidate sets with a beauty score. For each category, we run a small-scale pilot CrowdFlower experiment to consolidate the editors' assessment taking into account the micro-workers feedback. This process led us to mark some of the \textit{Test Question} with two contiguous scores. After this validation step, we identified the set of 100 images with the highest agreement that belongs to the full range of grades.

\subsection {Results} \label{sec:crowdflower:results}

We run a separate job for each topical category. Table~\ref{tab:exp-stats} summarizes the number of units annotated, judgments submitted, distinct participants, and the average accuracy (trust) on \textit{Test Questions} of the contributors. Each unit can receive more than 5 independent judgments; in the case of \textit{nature} we collected $20\%$ more judgments than for the other categories. On average, more than 140 contributors geographically distributed in 13 countries and characterized by a high level of trustworthiness participated to each experiment.  

\begin{table}[t!]
\begin{center}
\begin{small}
\begin{tabular}{| l | c | c | c | c |  c |}
\hline
 & Units &  Judgments & Workers & Countries & Trust \\ \hline
\textbf{people} & 2500 & 12725	 & 141 & 13  & 0.843 \\
\textbf{nature} & 2500 & 15054 & 178 & 14 & 0.841 \\
\textbf{animals} & 2500 & 13269 & 117 & 13  & 0.80 \\
\textbf{urban} & 2500 & 13213 & 111 & 13  & 0.839 \\ \hline
\end{tabular}
\end{small}
\end{center}
\vspace{-10pt}
\caption{\small General statistics on the crowdsourcing experiment}
\vspace{-1.5em}
\label{tab:exp-stats}
\end{table}%

\vspace{4pt} \noindent \textbf{Inter-rater agreement.} To assess the quality of the collected data, we measure the level of agreement between annotators. Table~\ref{tab:agreement} shows a set of standard measures to evaluate the inter-rater consistency. \textit{Matching\%} is the percentage of matching judgments per item. Across categories the agreement is solid, with an average of $70\%$. However, the ratio of matching grades does not capture entirely the extent to which agreement emerges. In fact, the task is inherently subjective and in some cases the quality of an image naturally converges to an intermediate level. We therefore compute the Fleiss' $K$, a statistical measure for assessing the reliability of the agreement between a fixed number of raters. Since Fleiss' $K$ is used to evaluate agreements on categorical ratings, it is not directly applicable to our task. We therefore binarize the task, and assign to each judgment either a  $Beautiful$ or $Not Beautiful$ label,  according to the score being respectively greater or lower than the median. Consistently, the Fleiss' $K$ shows a fair level of agreement. To further evaluate inter-participant consistency we computed the Cronbach's $\alpha$ that has been extensively adopted in the context of assessing inter-rater agreement on aesthetics tasks~\cite{redi2014beauty}. For all categories, the Cronbach's coefficient lies in the interval $0.7\leq\alpha<0.9$ that is commonly defined as a $Good$ level of consistency.

\begin{table}[t!]
\begin{center}
\begin{small}
\begin{tabular}{| l | c | c | c |}
\hline
 & Matching\% & Fleiss' $K$ & Cronbach's $\alpha$ \\ \hline

	\textbf{people} & 68.82  &  0.38 & 0.74 \\
	\textbf{nature} & 72.65  & 0.27 & 0.71\\
	\textbf{animals} & 69.37 &  0.35 & 0.8  \\
	\textbf{urban} & 73.13 & 0.38 & 0.8  \\ \hline
\end{tabular}
\end{small}
\end{center}
\vspace{-10pt}
\caption{\small Measures of judgment agreement}
\vspace{-0.5em}
\label{tab:agreement}
\end{table}%

\begin{figure}[t!]
\begin{center}
\includegraphics[width=0.8\columnwidth]{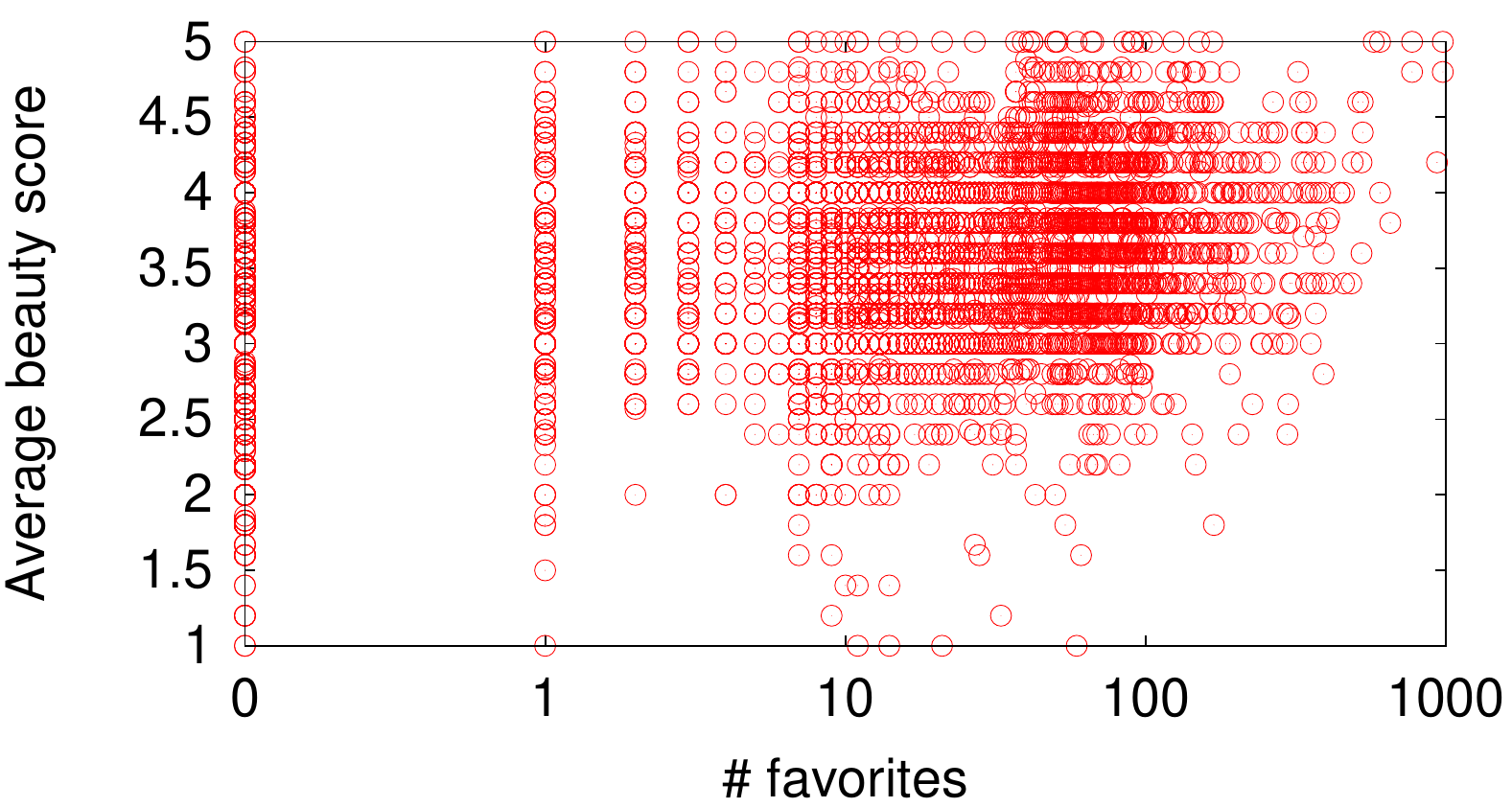}
\vspace{-8pt}
\caption{\small Relation between popularity (number of favorites) and crowdsourced beauty scores for 10,800 Flickr pictures.}
\vspace{-1.5em}
\label{fig:scatterplot}
\end{center}
\end{figure}

\vspace{4pt} \noindent \textbf{Beauty judgements.} The Spearman correlation $\rho$ between the number of favorites and the average beauty score is $0.43$. Although the correlation is substantial, the variability of perceived beauty for each popularity value is very high. In Figure~\ref{fig:scatterplot} we plot the beauty score against the number of favorites, for each photo. Zero-popularity images span the whole aesthetics judgment scale, from 1 to 5, and most popularity levels have photos within the $[2.5,5]$ beauty range. Very low scores (1,2) are rare. This picture confirms our initial motivation as it shows instances of unpopular yet beautiful photos, as well as a good portion of very popular photos with average or low quality.

\begin{figure}[t!]
\begin{center}
\includegraphics[width=0.8\columnwidth]{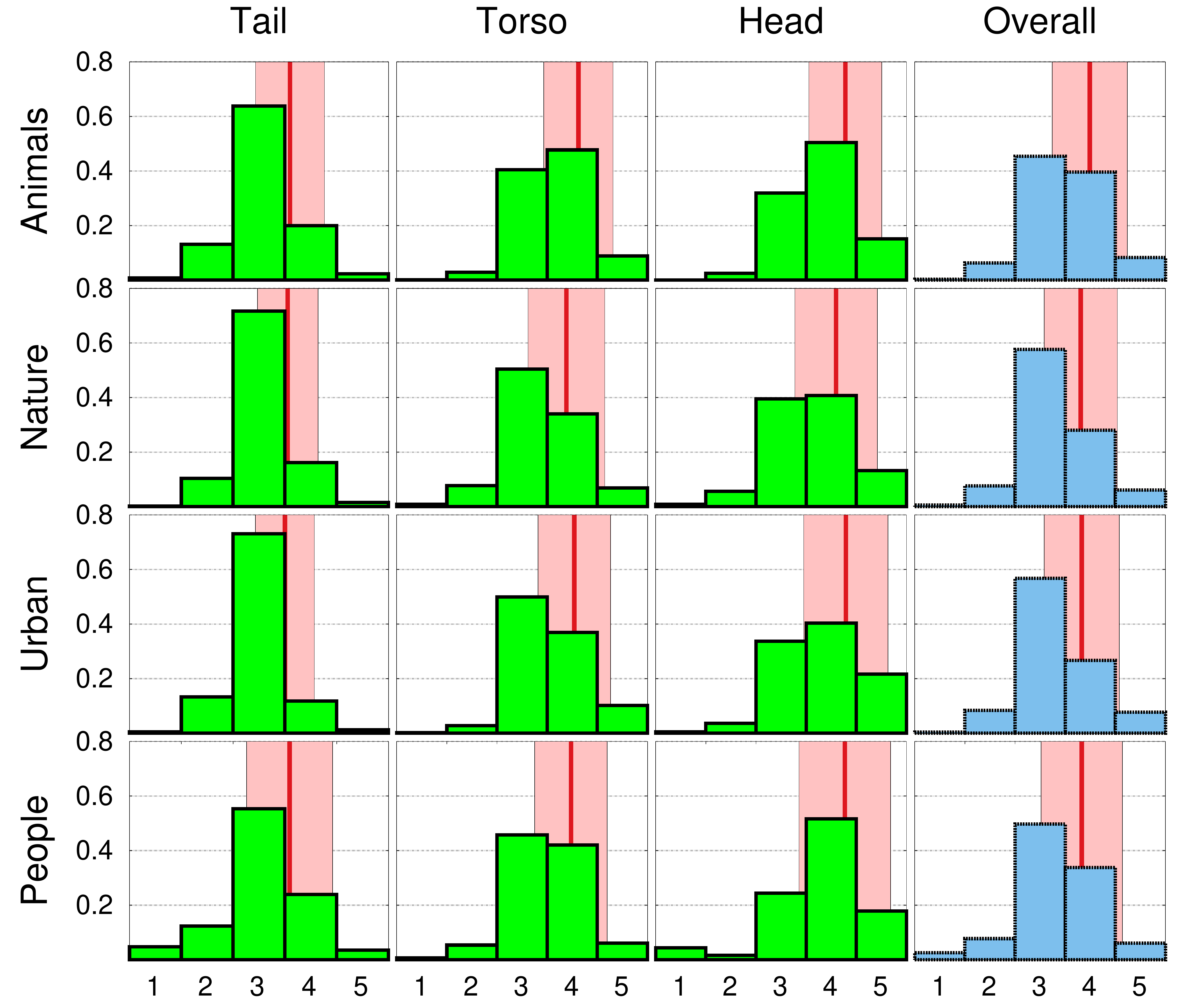}
\vspace{-8pt}
\caption{\small Distribution of ratings across categories and popularity buckets. The red lines and their surrounding areas represent the average and standard deviation.}
\vspace{-1.5em}
\label{tab:crowdflower_results}
\end{center}
\end{figure}

Results on the distribution of judgments across categories and popularity buckets are summarized in Figure~\ref{tab:crowdflower_results}. As expected, the $high$ bucket shows the highest average score followed by the $medium$ and the $low$. With the exception of the $people$ category, the standard deviation follows the same trend: higher popularity corresponds to higher disagreement. This might be due to the fact that viewers are likely to largely agree on objective elements that make an image non-appealing, such as technical flaws (e.g., bad focus) but on the other hand they might not agree on what makes an image exceptionally beautiful, which can be a more subjective characteristic. Given that the more a photo is popular the more it tends to be appealing, this phenomenon can partly explain the inconsistent agreement level among popularity buckets. Across categories we observe that $animals$ images have the highest average quality perception ($3.49\pm0.75$) while the remaining categories show a mean around $3.31$. 

\section{Image Aesthetics}\label{sec:aesthetics}

Having collected a ground truth of crowdsourced beauty judgements, we now design a computational aesthetic framework to surface beautiful, unpopular pictures. Our method is based on regressed \textit{compositional features}, namely visual features that are specifically designed to describe how much an image fulfills standard photographic rules.
We design our framework as follows:

\vspace{4pt}
\noindent \textbf{Visual Features.} We design a set of visual features to expose image photographic properties. More specifically, we compose a 47-dimensional feature vector with the following descriptors: 
\begin{itemize}
\item \textit{Color Features.}  Color patterns are important cues to understand the aesthetic and affective value of a picture. First, we compute a \textit{Contrast} metric, that provides information about the distinguishability of colors based on the magnitude of the average luminance: 
\begin{equation}
Contrast=\frac{Y_{max}-Y_{min}}{\bar{Y}}
\end{equation}
where $Y_{max},Y_{min},\bar{Y}$ correspond respectively to maximum, minimum, and average of the luminance channel.

We then extract the average of the \textit{Hue, Saturation, Brightness (H,S,V)} channels, computed both on the whole image and on the inner quadrant resulting after a 3x3 division of the image, similar to previous approaches~\cite{datta}. By combining average Saturation ($\bar{S}$) and Brightness ($\bar{V}$) values, we also extract  three  indicators of emotional dimensions, \textit{Pleasure, Arousal and Dominance}, as suggested by previous work on affective image analysis~\cite{emotions}:
\begin{equation}
\begin{split}
Pleasure &=   0.69 \bar{V} +0.22 \bar{S} \\
Arousal &=  -  0.31 \bar{V} +0.60 \bar{S} \\
Dominance &=  0.76 \bar{V} +0.32 \bar{S} 
\end{split}
\end{equation}
After quantizing the HSV values, we also collect the occurrences of 12 Hue bins, 5 Saturation bins, and 3 Brightness bins in the HSV \textit{Itten Color Histograms}. Finally, we compute \textit{Itten Color Contrasts}, i.e. the standard deviation of H, S and V Itten Color Histograms~\cite{emotions}. 
\item \textit{Spatial Arrangement Features.} Spatial arrangement of objects, shapes and people plays a key role in the shooting of good photographs~\cite{freeman2007photographer}. To analyze the spatial layout in the scene, first, we resize the image to a squared matrix $\mathbf{I_{ij}}$, and we compute a \textit{Symmetry} descriptor based on the difference of the Histograms of Oriented Gradients (HOG)~\cite{dalal2005histograms} between the left half of the image and its flipped right half:
\begin{equation}
Symmetry=||\Phi(\mathbf{I^l})-\Phi(\mathbf{(I\cdot J)^r})||_2,
\end{equation}
where $\Phi$ is the HOG operation, $\mathbf{I^l}$ is the left half of the image, and $\mathbf{(I\cdot J)^r}$ is the flipped right half of the image, being $\mathbf{J}$ the anti diagonal identity matrix that imposes the left-right flipping of the columns in $\mathbf{I_{ij}}$. We also consider the \textit{Rule of Thirds}, a photographic guideline stating that the important compositional elements of a picture should lie on four ideal lines (two horizontal, two vertical) that divide it into nine equal parts (the thirds). To model it, from the resized image $\mathbf{I_{ij}}$, we compute the a saliency matrix~\cite{hou2007saliency}, exposing the image regions that are more likely to grasp the attention of the human eye. We then analyze the distribution of the salient zones across the image thirds by retaining the average saliency value for each third subregion. 
\item \textit{Texture Features.}  We describe the overall complexity and homogeneity of an image by computing the Haralick's features~\cite{haralick1979statistical}, namely the \textit{Entropy, Energy, Homogeneity, Contrast} of the Gray-Level Co-occurrence Matrices.
\end{itemize}
\textbf{Groundtruth.} We use our crowdsourced groundtruth as the main source of knowledge for our supervised framework. Since topic-specific aesthetic models have been shown to perform better than general frameworks~\cite{luo2011content}, we keep the division of the ground truth into semantic categories (\textit{people}, \textit{urban}, \textit{animals}, \textit{nature}), and  learn a separate, topic-specific aesthetic model for each category.

\vspace{4pt} \noindent \textbf{Learning Framework.} We train category-specific models using \textit{Partial Least Squares Regression (PLSR)}, a very effecive prediction framework for visual pattern analysis \cite{mcintosh1996spatial}. For each semantic category, PLSR learns a set of \textit{regression coefficients}, one per dimension of the visual feature vector, by combining principles of least-squares regression and principal component analysis. Each category-specific group of regression coefficients constitutes a separate aesthetic model.

\vspace{4pt} \noindent \textbf{Prediction and Surfacing.} We apply the models to automatically assess the aesthetic value of new, unseen images (i.e., images that do not belong to the training set). To do so, we use the regression coefficients in a linear combination with the features of each image, thus obtaining  the predicted aesthetic score for that image.

\vspace{4pt} We use our aesthetic models for two types of experiments. First, to study the performance of our framework against similar approaches, we run a small-scale experiment where the task is to \textit{predict} the aesthetic scores of the crowdsourced groundtruth. We then apply the aesthetic models to \textit{rank} a very large set of images in terms of  beauty, with the aim of surfacing the most appealing non-popular pictures.

\section{Beauty Prediction from and for the Crowd }\label{sec:prediction}

\begin{table}[t!]
\resizebox{\linewidth}{!}{
\begin{tabular}{ | l | c | c |c | c|} 
\hline
	&CrowdBeauty & MIT popularity & TraditionalBeauty & Random \  \\ \hline
	\textbf{animals} & 0.54 & 0.37 & 0.251 &  0.001
	\\ \hline
	\textbf{urban}  & 0.46 & 0.27 & 0.12 & 0.003
	\\ \hline
	\textbf{nature}  & 0.34 & 0.29 & 0.11 & -0.003
	\\ \hline
	\textbf{people}  & 0.42 & 0.31 & 0.27 & -0.008 
	 \\ \hline
\end{tabular}}
\caption{\small Spearman correlation between the crowdsourced beauty judgments and the scores given by different methods on the images of the test set.}
\vspace{-1.5em}
\label{tab:correlations}
\end{table}

To test the power of our aesthetics predictor, we run a small-scale experiment on the crowd-sourced dataset. We look at how much the aesthetic scores assigned by our framework correlate with the actual beauty scores assigned by the workers, and evaluate the performance of our algorithm against other ranking strategies.

\vspace{5pt} \noindent \textbf{Baselines.} We compare our method with the following two baselines:

\vspace{2pt}
\noindent \textit{Popularity Predictor:} What if a popularity predictor was enough to assess image beauty? To check that, we compare our algorithm with an established content-based image popularity predictor. For each picture in our ground truth, we query the MIT popularity API\footnote{\scriptsize \url{http://popularity.csail.mit.edu}}, a recently proposed framework that automatically predicts image popularity scores (in terms of normalized view count) score given visual cues, such as colors and deep learning features~\cite{khosla14image}. 

\vspace{2pt}
\noindent \textit{Traditional Aesthetic Predictor:} What if existing aesthetic frameworks were general enough to assess crowdsourced beauty? As mentioned in $\S$\ref{sec:aesthetics}, our models are specifically trained on the crowdsourced dataset, i.e., a groundtruth of images generated and voted by average users. On the other hand, existing aesthetic predictors are generally trained on semi-professional images evaluated by professional photographers. To justify our dataset collection effort, we show how a classifier trained on traditional aesthetic datasets performs in comparison with our method. We design this baseline with the same structure and features as our proposed method, but, instead of using our crowdsourced ground truth, we train on the AVA dataset~\cite{murray2012ava}. Similar to our method, we build one category-specific model for each semantic category. This is achieved by training each category-specific model with the subset of AVA pictures in the corresponding category. We infer the category according to tags attached to each image, as proposed for many topic-specific aesthetic models~\cite{luo08photo,obrador09role}.

\vspace{3pt} \noindent \textbf{Experimental Setup.}
To evaluate our framework, for each  semantic category we retain 800 images for test and the rest for training. For training, we use images from all the 3 popularity ranges (tail, torso, head). For test, we consider non-popular images only, as our main purpose is to detect ``hidden'' beautiful pictures with low number of favorites. For both training and test, we use the total of 47 visual features, that are reduced to 15 components by the PLSR algorithm.

We then score the images in the test set using the output of our framework, the MIT popularity scores, the output of the traditional aesthetic classifier, and a random baseline. Next, we evaluate the performance of the three algorithms in terms of Spearman Correlation Coefficient between the scores predicted on the test set by each model, and the actual votes from the crowd. This metric gauges the ability of each model to replicate the human aesthetic preferences on non-popular Flickr images.

\begin{figure}[t!]
\begin{center}
\includegraphics[width=0.35\textwidth]{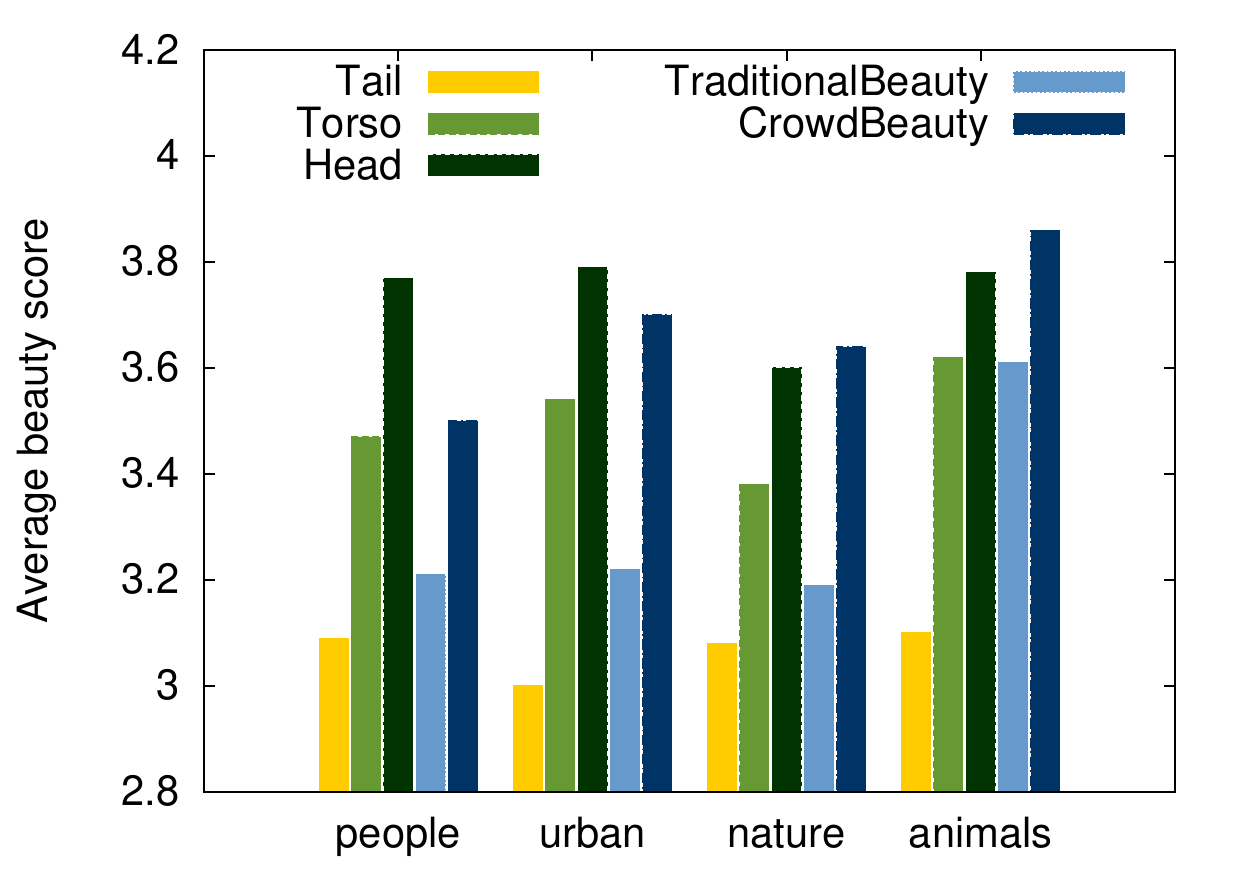}
\vspace{-10pt}
\caption{Average crowdsourced beauty score photos in different popularity buckets and for photos surfaced by the aesthetics predictors.}
\vspace{-1.5em}
\label{fig:results-retraining}
\end{center}
\end{figure}

\begin{figure*}[t!]
\begin{center}
      \includegraphics[width=.8\textwidth]{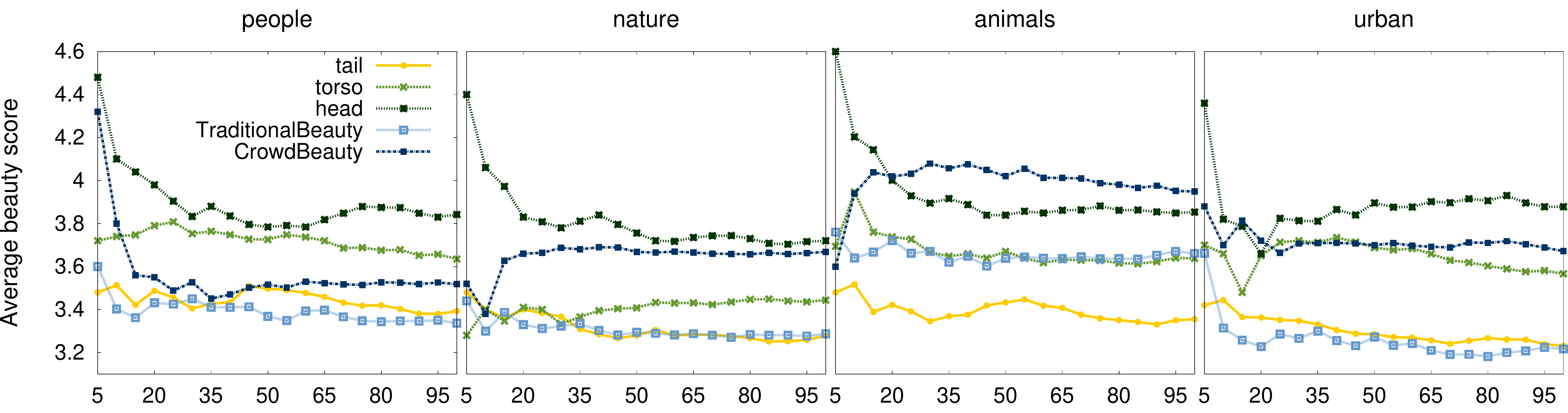}
		\vspace{-8pt}
    \caption{\small Average beauty of the top $n$ pictures ranked by popularity (in tail, torso, and head buckets) and by the predicted beauty scores.}
		\vspace{-1.5em}
    \label{fig:beauty_at_n}
		\end{center}
\end{figure*}

\vspace{5pt} \noindent \textbf{Experimental Results.}
The correlation between the beauty scores assigned by the micro-workers on the test set and our proposed algorithm (\textit{CrowdBeauty} in the following) is substantially high for all categories, as shown in Table~\ref{tab:correlations}. In particular, the most predictable class is the \textit{animals} category, followed by \textit{urban}. The higher performance in these two cases might be due to the smaller range of poses and compositional layouts available to the photographer when shooting pictures of subjects belonging to these particular categories. As expected, the results of the random approach are completely uncorrelated from the beauty scores.
 For all semantic categories, we see that our method outperforms both the popularity predictor (\textit{MIT Popularity}) and the traditional aesthetic classifier (\textit{TraditionalBeauty}), showing the usefulness of building a dedicated ground truth and aesthetic classifier to score non-popular web images. 

\section{Surfacing Beautiful Hidden Photos}\label{sec:surfacing}

Having provided some evidence about the effectiveness of our approach, we apply it in a more realistic scenario where the goal is to surface beautiful images from a large number of non-popular Flickr pictures. 

To do so, we compute the features described in $\S$\ref{sec:aesthetics} on all the 9M images of the large-scale categorized dataset of creative commons Flickr images in our dataset. We apply the category-specific model on the pictures in each topical category separately and rank the pictures by their predicted aesthetics scores. For the sake of comparison, we repeat the same procedure with the traditional aesthetic models (\textit{TraditionalBeauty}) used as baseline in $\S$\ref{sec:prediction}, and rank them in terms of the predicted beauty scores. We do not consider here the MIT Popularity baseline as its scores can only be retrieved via API with a certain request delay, which it is not practical for a very large set of images.

To quantify how appealing the images surfaced with our approach are, we implemented an additional crowdsourcing experiment in which images with different popularity levels are evaluated against the top-ranked images according to our models and the traditional aesthetic model. 
We replicated the same experimental settings described in Section~\ref{sec:crowdflower} and we used a corpus composed by 200, 200 and 100 images from the tail, torso and head of the popularity distribution respectively, and we added the top 200 images from the \textit{TraditionalBeauty} and \textit{CrowdBeauty} rankings. For consistency, we maintained the same proportion of items per class we used in the previous experiments, but with a smaller sample that focuses only on the top ranked beautiful images. 

Figure~\ref{fig:results-retraining} shows the average beauty score for each category and bucket combination. Consistently across categories, the perceived beauty of the \textit{CrowdBeauty} images is comparable to the most favorited photos. In fact, for \textit{nature} and \textit{animals} we observe an average increment of $0.9\%$ and $1.3\%$ with respect to most popular items and for \textit{urban} and \textit{people} a decrease of $2\%$ and $7\%$, respectively. With the exception of \textit{people}, the median of the perceived beauty score goes up from 3 to 4 when \textit{CrowdBeauty} is adopted against \textit{TraditionalBeauty}. This behavior confirms how important the training of an aesthetic predictor with a reliable ground truth is for this task. 

The study of the average behavior of the beauty predictors does not show what happens if we consider only the head of the rank. For some applications this could be relevant, e.g., recommender systems suggest the top $n$ most relevant items for a user. To this extent it is interesting to evaluate the perceived beauty of the topmost images. Figure~\ref{fig:beauty_at_n} shows for each category how the average beauty score varies at cutoffs $n\in[5,100]$. Highly popular items have a consistent behavior across categories where items at the top of the rank are perceived as very appealing and the quality drops and stabilizes quickly after $n=20$. In general, after an initial variation, \textit{CrowdBeauty} stabilizes above the tail, torso and \textit{TraditionalBeauty} curves. If \textit{urban} is almost stable for all the cutoffs, \textit{nature} and \textit{animals} start with lower quality items and rapidly jump to higher values. A different case is the \textit{people} category where the top ten images have a very high score and then they drop after n=20.

Some examples of highly ranked images surfaced by our algorithm alongside with the least and most favorited pictures are shown in Table~\ref{tab:examples}.

\begin{table}[t]
\begin{center}
\begin{tabular}{cccc}
 & Tail & Head & CrowdBeauty \\

\begin{turn}{90} \hspace{4mm}\textbf{Animals}\end{turn} & \includegraphics[width = 0.12\textwidth]{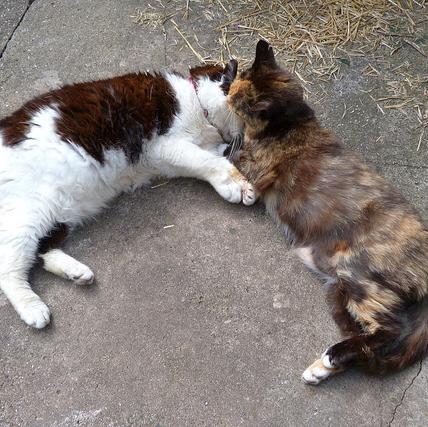} & \includegraphics[width = 0.12\textwidth]{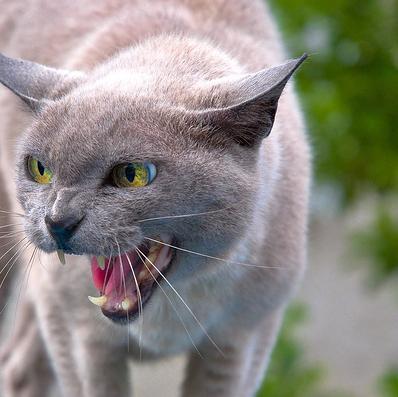}  & \includegraphics[width = 0.12\textwidth]{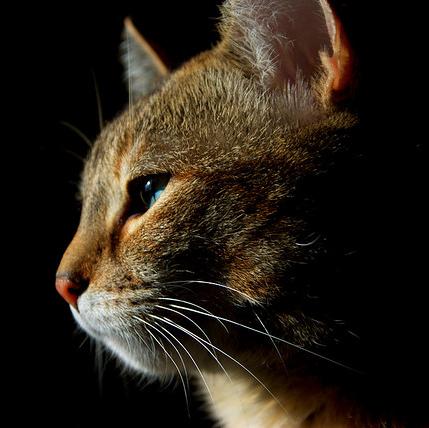}\\

\begin{turn}{90} \hspace{4.5mm}\textbf{People}\end{turn} & \includegraphics[width = 0.12\textwidth]{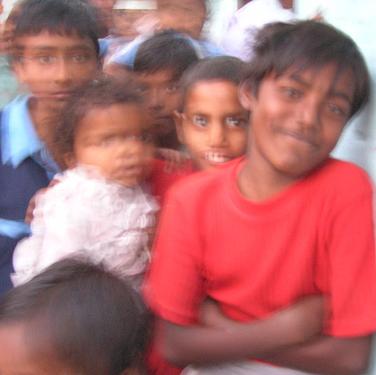} & \includegraphics[width = 0.12\textwidth]{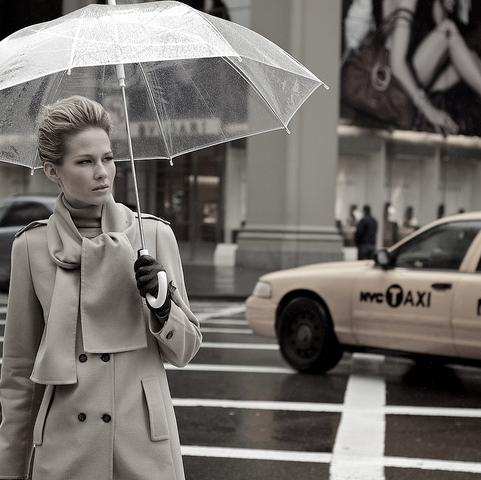} & \includegraphics[width = 0.12\textwidth]{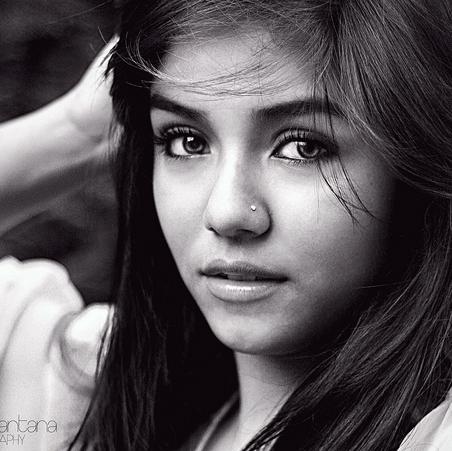} \\

\begin{turn}{90} \hspace{4.5mm}\textbf{Urban}\end{turn} & \includegraphics[width = 0.12\textwidth]{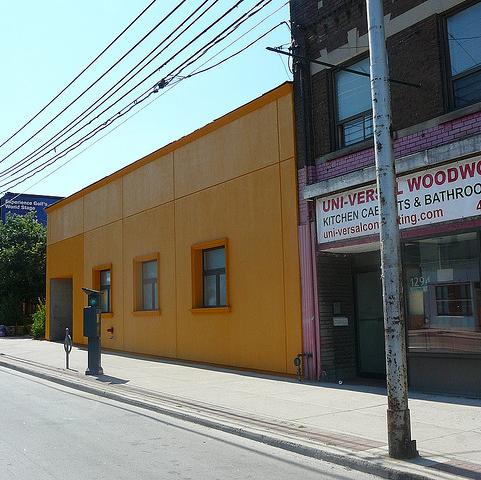} & \includegraphics[width = 0.12\textwidth]{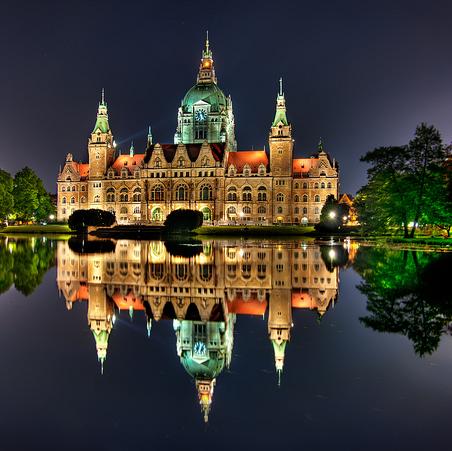} & \includegraphics[width = 0.12\textwidth]{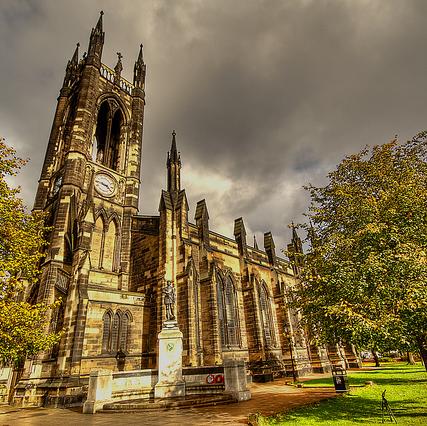} \\

\begin{turn}{90} \hspace{4.5mm}\textbf{Nature}\end{turn} & \includegraphics[width = 0.12\textwidth]{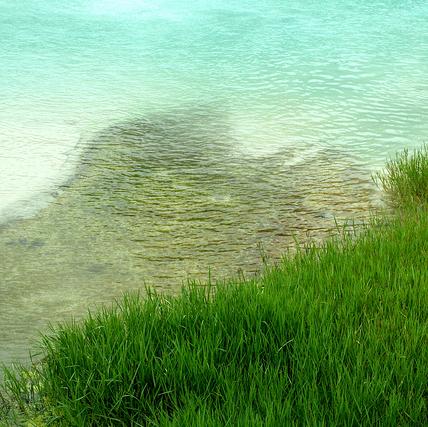} & \includegraphics[width = 0.12\textwidth]{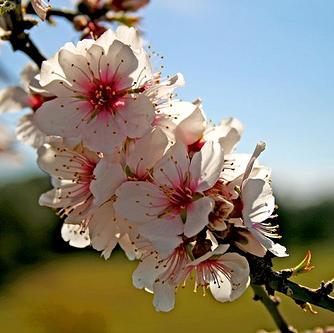} & \includegraphics[width = 0.12\textwidth]{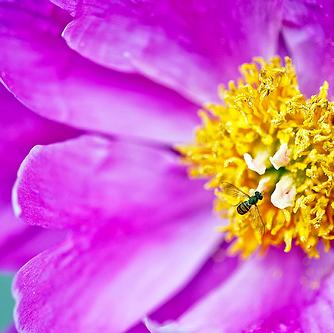} \\

\end{tabular}
\caption{\small Samples of images from \textit{tail} and \textit{head} popularity buckets, compared to the images surfaced by our approach.}\label{tab:examples}
\end{center}
\vspace{-20pt}
\end{table}
  
\section{Discussion and Conclusions}\label{sec:conclusions}

\noindent \textbf{Applications and future work.} The ability to rank by aesthetic appeal images that are nearly indistinguishable in terms of the user feedback by aesthetic value has immediate applications. First, it promotes the \textit{democratization} of photo sharing platforms, creating an opportunity to balance the visibility of popular and beautiful photos with those that are as beautiful but with less social exposure. As a proof-of-concept, we envision a new Flickr \textit{Beauty Explorer} page that surfaces the most beautiful yet unpopular photos of the month to complement the classic Flickr \textit{Explorer} that contains photos with very high social feedback. Our method can be used to bring valuable but unengaged users into the active core of the community by canalizing other people's attention towards them. An extension to this work could be to use the aggregation of photo quality over users to spot hidden \textit{talents} and devise incentive mechanisms to prevent them to churn. Furthermore, our method increases the payoff of the service provider by uncovering valuable content, exploitable for promotion, advertising, mashup, or any other commercial service, that would have been nearly useless otherwise. Also it would be interesting to study the effect of aesthetic reranking on the \textit{head} of the popularity distribution, or on images relevant to a specific query.


\vspace{5pt} \noindent \textbf{Limitations.} Our approach comes with a few limitations, mainly introduced by the computer vision method we use. 

\begin{figure}[t!]
    \subfloat[Animals\label{fig:bad_a}]{%
     \includegraphics[width = 0.15\textwidth]{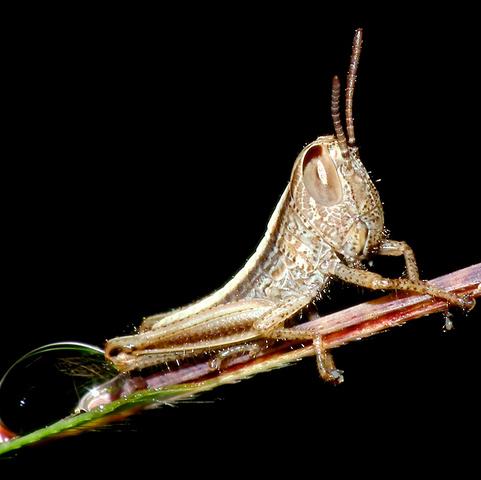}
    }
    \hfill
    \subfloat[Urban\label{fig:bad_b}]{%
   \includegraphics[width = 0.15\textwidth]{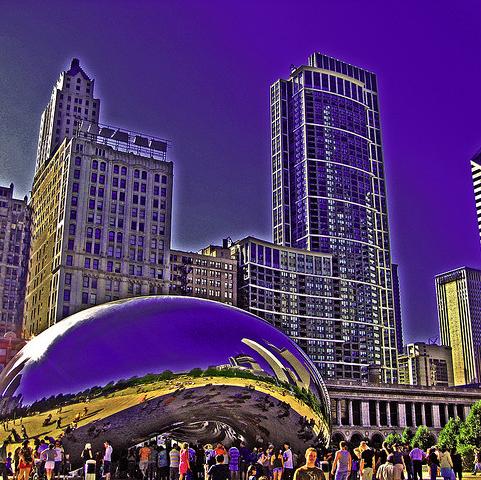}
    }
  \hfill
    \subfloat[People\label{fig:bad_c}]{%
       \includegraphics[width = 0.15\textwidth]{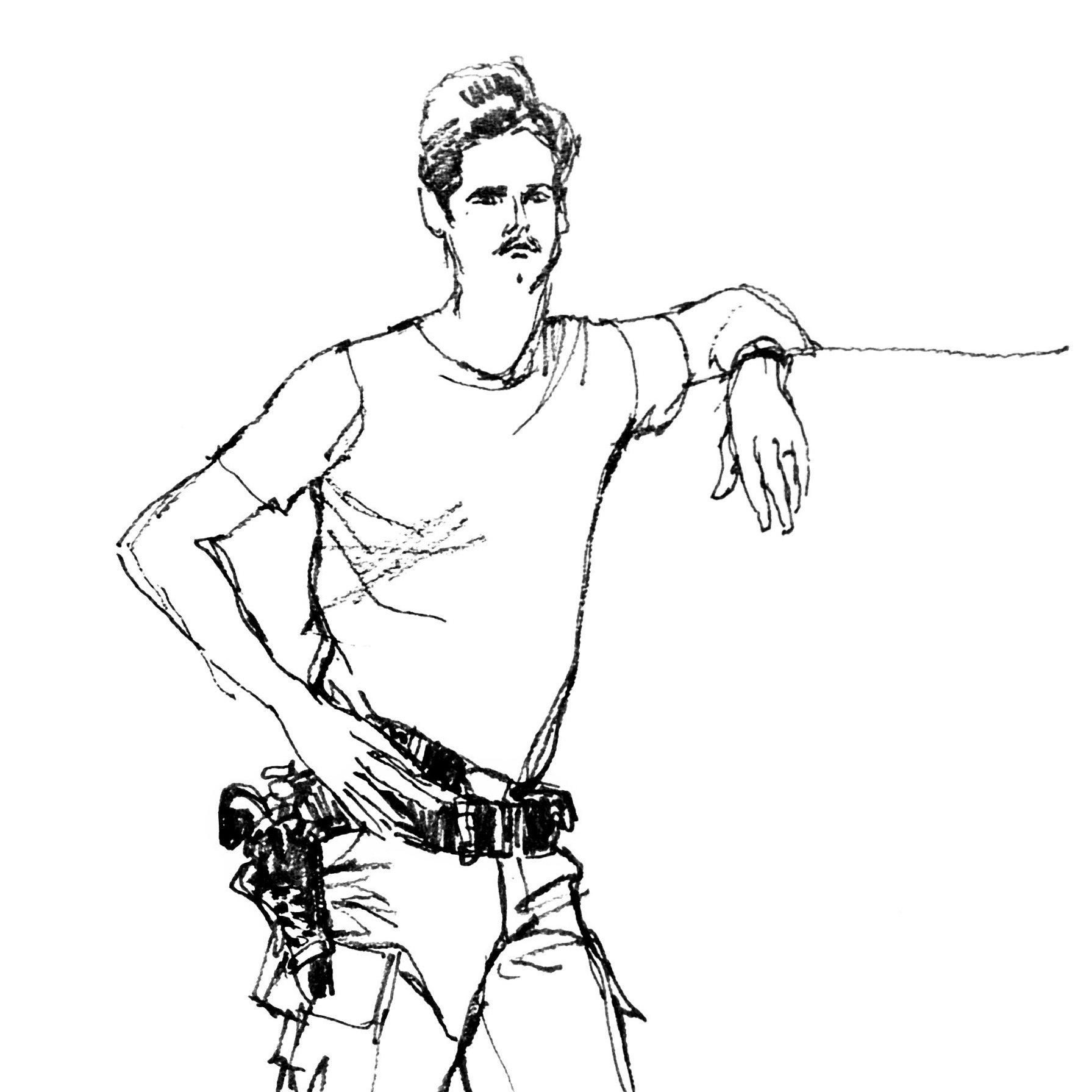}
    }
\caption{\small Examples of biases in surfaced pictures.}
\vspace{-1.5em}
    \label{fig:bad_examples}
  \end{figure}

First, although \textit{machine-tags} have a very high accuracy, they sometimes recognize objects even when they are simply drawn or sketched, and attach semantic tags to non-photographic images, e.g., clipart (see Figure~\ref{fig:bad_c}). Non-photographic images have their own aesthetic rules that differ substantially from photographs, and photo aesthetic predictors typically give erroneous predictions on non-photographic images. While in this work we manually removed some non-photographic images from our corpus to allow the model to smoothly learn photographic aesthetic rules, an automatic pre-filtering bassed on non-photographic image detectors would be advisable~\cite{ng2007lessons}.

Second, despite the high quality of the surfaced photos, some top-ranked \textit{animals} and \textit{nature} images receive lower scores than some lower-ranked ones. This behavior is due to biases in the learning framework: some of the top-rated images for \textit{animals} and \textit{nature} are extremely contrasted pictures (see Figure~\ref{fig:bad_a}) thus the model wrongly over-weights the contrast features. Similarly, some of the surfaced \textit{urban} pictures show strong presence of contrast/median filtering, such as the example in Figure~\ref{fig:bad_b}.

Last, our method is less effective in surfacing good \textit{people} images. Often highly rated pictures of people show black and white color palette, thus biasing the aesthetic model. From a broader perspective, pictures of people are different in nature from other image types. Faces grasp human attention more than other subjects~\cite{bakhshi14faces}: face perception is one of the most developed human skills~\cite{haxby2000distributed}, and that we have  brain sub-networks dedicated to face processing~\cite{freiwald2014neurons}. Moreover, when shooting photos of people, photographers need to capture much more than the traits of the mere subject: people come with their emotions, stories, and lifestyles. Portrait photography is indeed a separate branch of traditional photography with dedicated books and compositional techniques~\cite{weiser1999phototherapy,child2008studio,hurter2007portrait}. The traditional compositional features that we use in our framework can only partially capture the essence of the aesthetics of portraits.

\vspace{5pt} \noindent \textbf{Concluding remarks.} The popularization of online broadcast communication media, the resulting information overload, and the consequent shrinkage of the attention span online have shaped the Social Web increasingly towards a frantic search for popularity, that many users yearn for. In this rampant race for fame that very few can win, the crowd often cannot see (and sometimes tramples on) some of the valuable gems that itself creates. To fix that in the context of photo sharing systems, we show that it is possible to apply computer vision techniques that spot beautiful images from the immense and often forgotten mass of pictures in the popularity tail. To do that, we show the necessity of using dedicated crowdsourced beauty judgements done by common people on common people's photos, in contrast to corpora of professional photos annotated by professionals. We hope that our work can be a cautionary tale about the importance of targeting content quality instead of popularity, not just limited to multimedia items but in social media at large.

\section*{Acknowledgments}
R. Schifanella was partially supported by the Yahoo FREP grant. We thank Dr. Judith Redi for her precious help and discussions.

\bibliographystyle{aaai}
\small
\bibliography{bibliography_short}  
\end{document}